\documentclass[letterpaper, 10 pt, doublespace]{IEEEtran}  

\IEEEoverridecommandlockouts

\usepackage{amsmath} 
\usepackage{amssymb} 
\usepackage{tabularx} 
\usepackage{fancyhdr}
\usepackage[ruled,vlined,linesnumbered]{algorithm2e}
\usepackage{amsthm}
\usepackage{graphicx}
\usepackage[space]{grffile}
\usepackage{cite}
\usepackage{wrapfig}
\usepackage{hyperref}
\usepackage{subfig}
\usepackage{url}
\usepackage[space]{grffile}

\expandafter\def\expandafter\normalsize\expandafter{%
    \normalsize
    \setlength\abovedisplayskip{4pt}
    \setlength\belowdisplayskip{4pt}
    \setlength\abovedisplayshortskip{4pt}
    \setlength\belowdisplayshortskip{4pt}
}

\title{\LARGE \bf
Optimal Torque Control of Permanent Magnet Synchronous Motors Using Adaptive Dynamic Programming
}

\author{
	\begin{tabular}[t]{c@{\extracolsep{10em}}c} 
		Ataollah Gogani Khiabani  & Ali Heydari \\
	\end{tabular}
}

\begin{document}

\maketitle
\pagenumbering{arabic}
\pagestyle{fancy}
\thispagestyle{fancy}
\fancyhead{}
\fancyhead[C]{This paper is a postprint of a paper submitted to and accepted for publication in
	IET Power Electronics and is subject to Institution of Engineering and Technology Copyright. The copy of record is
	available at the IET Digital Library}

\begin{abstract}
	In this study, a new approach based on adaptive dynamic programming (ADP) is proposed to control permanent magnet synchronous motors (PMSMs). The objective of this paper is to control the torque and consequently the speed of a PMSM when an unknown load torque is applied to it. The proposed controller achieves a fast transient response, low ripples and small steady-state error. The control algorithm uses two neural networks, called critic and actor. The former is utilized to evaluate the cost and the latter is used to generate control signals. The training is done once offline and the calculated optimal weights of actor network are used in online control to achieve fast and accurate torque control of PMSMs. This algorithm is compared with field oriented control (FOC) and direct torque control based on space vector modulation (DTC-SVM). Simulations and experimental results show that the proposed algorithm provides desirable results under both accurate and uncertain modeled dynamics. Although the performance of FOC method is comparable with ADP under nominal conditions, the torque and speed response of ADP is better than FOC under realistic scenarios, that is, when parameter uncertainties exist.
\end{abstract}

\textit{Index Terms-} Adaptive Dynamic Programming (ADP), Permanent magnet synchronous motor (PMSM), Direct torque control, Parameter uncertainty compensation 

\maketitle

\section{Introduction}
In recent years, there has been an increasing attention to permanent magnet synchronous motors (PMSMs) because of their undeniable advantages, such as high power density, high torque to inertia ratio and reliability \cite{melfi2009induction, vaez2018control}. The performance of a PMSM highly depends on the quality of its control scheme. In many applications such as electric vehicles (EVs) and robotics, it is necessary to achieve precise control of the motor. The control should be satisfactory in the presence of system parameter uncertainties as well as external disturbances.

Field oriented control (FOC) and direct torque control (DTC) are the two main control approaches of alternate current (AC) servo drives. A typical FOC scheme consists of two inner current loops and one outer speed loop. Proportional-integral (PI) controllers are commonly used to regulate the motor currents. In order to avoid large overshoots, the bandwidth of these current controllers is limited, which leads to the slow dynamic response of the motor, \cite{liu2019model}. Furthermore, the PI gains play a crucial role in the steady state behavior of the motor under FOC, therefore fine tuning of the gains is necessary \cite{blaschke1972principle, lim2013comparative}. In order to control a PMSM through power inverters, typically space vector modulation (SVM) is used to realize the appropriate voltage, which is applied to the motor. This method can provide voltage vectors with adjustable amplitude and phase, \cite{vaez2018control}.

DTC, on the other hand, utilizes another approach for the control. In classical DTC, based on two hysteresis comparators and a switching table, a suitable voltage vector is applied on the motor for the whole sampling time \cite{zhong1997analysis}. Although DTC provides a faster dynamic response compared to FOC, it has major disadvantages such as increased torque and stator flux ripples, \cite{liu2019model}. Furthermore, DTC requires high sampling frequency for digital implementation of the controller, which in turn, demands more powerful digital signal processors (DSPs) and increases the cost of implementation \cite{niu2015comparative}. In recent years, there has been many studies to address the disadvantages of the classical DTC, \cite{vafaie2015new}. One of these approaches utilizes DTC with SVM. Unlike DTC, which uses one of the available fixed voltage vectors, with a fixed amplitude and phase, for the duration of the control cycle, any arbitrary voltage vector can be generated and applied to the motor, when DTC is augmented with SVM (i.e., DTC-SVM). This voltage vector, which has an adjustable amplitude and phase, is generated using multiple vectors during the sampling time. This approach leads to a reduction in torque and flux ripples \cite{lai2001new, kenny2003stator, lascu2004combining}.

Recently, there has been research on more advanced control approaches for PMSMs, such as sliding mode control (SMC) \cite{dominguez2013digital, xu2018sliding, liu2017torque}, model predictive control (MPC) \cite{liu2019model, chen2019modified, sandre2019modified, bozorgi2017model, xu2019robust}, adaptive control \cite{nguyen2018model, jung2014adaptive}, fuzzy control \cite{chaoui2011adaptive, feng2016closed}, and neural network control \cite{ei2010hybrid, yi2003implementation}. Each of these controllers offers some improvement in the performance of PMSMs. In \cite{dominguez2013digital}, a digital sliding mode controller is designed to track the desired motor speed, and a digital observer is utilized to estimate the rotor position and velocity. The authors in \cite{liu2019model} have designed a model predictive controller for direct speed control of a PMSM. Although the results show great improvement, one disadvantage of this approach is the need for load torque observer. It may be noted that the load on the motor is typically unknown and varying, in reality. Therefore, it is desired to develop a controller with good robustness toward an uncertain load. In \cite{chen2019modified}, the authors have pursued an MPC-based approach in which, instead of applying one voltage vector during the control cycle, they apply two voltage vectors and calculate their respective duty cycle, i.e. the percentage of sampling period for which, a specific voltage vector is applied. Since in this approach, the torque and flux need to be estimated, this approach is not relatively robust to parameter uncertainties. An adaptive speed controller is proposed in \cite{nguyen2018model} in which, the controller is divided into two parts, one for stabilization of the error dynamics and the other one for dealing with parameter uncertainties. An artificial neural network (ANN) is used in \cite{yi2003implementation} to control the speed of a PMSM. In this approach, the inner current loops remain like classical FOC, however instead of a PI controller for the outer loop speed controller, an ANN is used to generate the reference current.

Recently, adaptive dynamic programming (ADP) has been used extensively in different applications to solve different optimal control problems \cite{kiumarsi2017optimal, sardarmehni2019sub, zhou2019neuro, khiabani2019design}. Usually, there are two neural networks in an ADP design. The first one is called the `critic', which approximates the value function (i.e. optimal cost-to-go) and the second neural network is called the `actor', which generates the optimal control based on the optimal value function and system states. ADP also has been used in the control of PMSMs in \cite{wang2018novel, li2019neural}. In \cite{wang2018novel}, a single artificial neural network based on ADP is designed which substitutes the outer loop PI speed controller in classical FOC. There are some disadvantages to this approach. First, the inner current loops are still FOC-base PI controllers, therefore the dynamic response of the control is not fast compared to other approaches such as DTC. Secondly, at each sampling time, there is a need for three consecutive speed error values, instead of only the error at that specific sampling time. In a simultaneous but independent research, the authors in \cite{li2019neural}, have designed a neural network controller which substitutes the inner loop PI current controllers in the classical FOC to deal with a decoupling inaccuracy issue of FOC. This controller has some relative robustness to parameter uncertainties.

The proposed approach in this research, for PMSM control, consists of two steps. The first step is done offline, such that based on a cost function and the system dynamics, the critic and actor network weights are calculated using value iteration (VI), \cite{heydari2014revisiting}. Then the actor weights are used in the online control to find the suitable control input in a feedback form. Therefore, the computational load in the online control stage is as low as evaluating a few algebraic functions (a feed forward of the actor). The control inputs are the voltages applied to the SVM block.

The contribution of this work is utilizing ADP for optimal control of PMSMs. This controller leads to a fast dynamic response as well as desirable steady state performance. The strength of this controller is observed more clearly when there are parameter uncertainties and load disturbances in the system. Moreover, after doing the offline training, the calculation load of this controller is as low as evaluating a polynomial function, which is extremely low compared to many other methods, including DTC or MPC. The low complexity of online implementation of this controller makes it a potential candidate for many applications. The proposed controller is compared with classical FOC and DTC-SVM, both in simulation and experiment, to show its superior performance.

As for organization of the paper, Section \ref{PMSM_model}, provides the dynamic equations of a PMSM. Afterwards, the optimal control using ADP in its general form is formulated in Section \ref{Optimal_control_ADP}. In Section \ref{ADP_PMSM}, the proposed ADP-based controller is simulated on a PMSM. The proposed controller is implemented on a physical prototype and the experimental results are provided in Section \ref{ADP_PMSM_experiment}, followed by concluding remarks in Section \ref{Conclusion}.
\vspace{-.1in}
\section{Dynamics of PMSM} \label{PMSM_model}

It is common to write the machine model of a PMSM in the \textit{dq} synchronous coordinates \cite{vaez2018control}. The parameters in \textit{dq} reference frame can be obtained from \textit{abc} parameters using Park-Clark transform as follows
\begin{equation}
\resizebox{.95\hsize}{!}
{$\begin{bmatrix} 	f_d \\ f_q \\ f_0 \end{bmatrix}
	=
	\frac{2}{3}\begin{bmatrix} \cos(\theta_e) & \cos(\theta_e-\frac{2\pi}{3}) & \cos(\theta_e+\frac{2\pi}{3}) \\
	-\sin(\theta_e) & -\sin(\theta_e-\frac{2\pi}{3}) & -\sin(\theta_e+\frac{2\pi}{3}) \\
	\frac{1}{2} & \frac{1}{2} & \frac{1}{2} \end{bmatrix}
	\begin{bmatrix} 	f_a \\ f_b \\ f_c \end{bmatrix}$},                      
\label{PK_transform}
\end{equation}
where $f$ can be a phase voltage, a phase current or a phase flux linkage. Also $\theta_e$ is the rotor \textit{electrical angle} between phase ``$a$'' of the stationary \textit{abc} reference frame and the ``\textit{d}-axis'' of the rotor reference frame and is calculated as
\begin{equation}
\theta_e=P\theta_m,
\label{electrical_angle}
\end{equation}
where $\theta_m$ is the rotor \textit{mechanical angle} and $P$ is the number of pole pairs. After transforming motor voltages and currents from \textit{abc} to \textit{dq} coordinates, the motor dynamic equations are obtained as
\begin{equation}
\begin{split}
\dot{\bar{x}}=\frac{d}{dt} 
\begin{bmatrix} 	i_d \\ i_q  \end{bmatrix}
&= 
\begin{bmatrix} 	L_d^{-1}(-R_si_d+L_qP\omega_mi_q )\\ L_q^{-1}(-R_si_q-L_dP\omega_mi_d-\lambda_mP\omega_m )   \end{bmatrix}\\
&+\begin{bmatrix} 	L_d^{-1} & 0\\ 0 & L_q^{-1}  \end{bmatrix}
\begin{bmatrix} 	v_d\\ v_q \end{bmatrix}
=\bar{f}(\bar{x})+\bar{g}\bar{u} \label{motor_dynamics}
\end{split}
\end{equation}
where $v_d$, $v_q$, and $i_d$, $i_q$ are the stator voltage and current in \textit{dq} reference frame. $R_s$ is the stator winding resistance, $L_d$, $L_q$ are stator winding inductance in \textit{dq} coordinates, $\omega_m$ is the rotor mechanical speed, and $\lambda_m$ is the magnetic flux linkage of the rotor permanent magnets. The torque balance equation of the motor is
\begin{equation}
\frac{d}{dt}\omega_m=\frac{1}{J}(\tau_{em}-\tau_f-\tau_L),
\label{torque_balance}
\end{equation}
where $J$ is the rotor and load inertia, $\tau_f$, and $\tau_L$ are friction and load torques, respectively. The motor electromagnetic torque, denoted with $\tau_{em}$, is given by 
\begin{equation}
\tau_{em}=\frac{3}{2}P\big((L_d-L_q)i_di_q+\lambda_mi_q\big),
\label{motor_torque}
\end{equation}
Finally, it may be mentioned that $\bar{x}$, denoting the state vector in (\ref{motor_dynamics}), is composed of $i_d$ and $i_q$. Moreover, the control vector is denoted with $\bar{u}$ and is composed of voltages $v_d$ and $v_q$.


\section{Optimal Control Using ADP} \label{Optimal_control_ADP}
\vspace{-.05in}
In this section, the optimal control problem is formulated and its solution using ADP is presented. Let the infinite-horizon cost function, subject to minimization, be given by
\begin{equation}
J=\sum_{k=0}^{\infty} \gamma^k\big(Q(x_k)+u_k^TRu_k\big),
\label{cost_function}
\end{equation}
where $x_k\in\mathbb{R}^n$ and $u_k\in\mathbb{R}^m$ are the system states with dimension $n$, and the control vector with dimension $m$, respectively. Moreover, $Q:\mathbb{R}^n \to \mathbb{R}_+$ penalizes the states and $R\in\mathbb{R}^{m\times m}$ penalizes the control. Furthermore, $\gamma \in (0,1]$ is the discount factor, which is used to ensure the boundedness of the cost. The discrete-time nonlinear dynamics is defined as
\begin{equation}
x_{k+1}=f(x_k)+g(x_k)u_k, k\in\{0,1,2,...\}, x(0)=x_0,
\label{nonlinear_dynamics}
\end{equation}
where $f:\mathbb{R}^n \to \mathbb{R}^n$ is a smooth function which represents the internal dynamics of the system and $g:\mathbb{R}^n \to \mathbb{R}^{n\times m}$, is the input gain function. The objective is to find the sequence of `optimal' control, denoted with $u_k^*, k\in\{0,1,2,...\}$ such that the cost function in (\ref{cost_function}) is minimized subject to system dynamics in (\ref{nonlinear_dynamics}). 

One can write the cost-to-go from current time to infinity, as a function of current state and future decisions, denoted with $V(.,.)$, as
\begin{equation}
V(x_k,\{u_h\}_{h={k}}^{\infty})=\sum_{h=k}^{\infty} \gamma^k\big(Q(x_h)+u_h^TRu_h\big).
\label{cost_to_go}
\end{equation}
It is possible to write the above equation in a recursive form as
\begin{equation}
\resizebox{.99\hsize}{!}
{$\begin{split}
	V(x_k,\{u_h\}_{h={k}}^{\infty})&= Q(x_k)+u_k^TRu_k+\gamma V(x_{k+1},\{u_h\}_{h={k+1}}^{\infty})\\
	& =Q(x_k)+u_k^TRu_k\\ &+\gamma V(f(x_k)+g(x_k)u_k,\{u_h\}_{h={k+1}}^{\infty}),\\
	\label{recursive_cost_to_go}
	\end{split}
	$}
\end{equation}
Function $V^*(x_k)$ is called the \textit{value function} which is the optimal cost-to-go from current state at current time to infinity, if the optimal control sequence is applied on the system. Considering the relation given by (\ref{recursive_cost_to_go}), one can find the value function and optimal control sequence based on Bellman principle of optimality \cite{kirk2012optimal} as follows
\begin{equation}
\begin{split}
V^*(x_k)&=inf_{\{u_h\}_{h={k}}^{\infty}}\big(V(x_k,\{u_h\}_{h={k}}^{\infty}) \big)\\
&=inf_{u_k\in\mathbb{R}^m}\Big(Q(x_k)+u_k^TRu_k \\ &+\gamma V^*\big(f(x_k)+g(x_k)u_k\big)\Big),\forall x_k\in \mathbb{R}^n,
\label{bellman_cost_to_go}
\end{split}
\end{equation}

\begin{equation}
\begin{split}
u^*(x_k)&=arginf_{u_k\in\mathbb{R}^m}\Big(Q(x_k)+u_k^TRu_k \\ &+\gamma V^*\big(f(x_k)+g(x_k)u_k\big)\Big),\forall x_k\in \mathbb{R}^n,
\label{bellman_control}
\end{split}
\end{equation}
which leads to 
\begin{equation}
\begin{split}
u^*(x)=-\frac{1}{2}\gamma R^{-1}g(x)^T\nabla V^*\big(f(x)+g(x)u^*(x)\big),\forall x\in \mathbb{R}^n,
\label{bellman_control_final}
\end{split}
\end{equation}
\begin{equation}
\begin{split}
V^*(x)&=Q(x)+u^*(x)^TRu^*(x)\\ &+\gamma V^*\big(f(x)+g(x)u^*(x)\big),\forall x\in \mathbb{R}^n,
\label{bellman_cost_to_go_final}
\end{split}
\end{equation}
where $\nabla V(x)$ is defined as $\partial V(x)/\partial x$. Therefore, theoretically the Bellman equation gives the solution to the optimal control problem. However, because of the curse of dimensionality \cite{kirk2012optimal}, it is not practically possible to solve this optimal control problem for most nonlinear systems. One approach of solving this optimal control problem is by the value iteration (VI)-based ADP. In this approach, the value function and optimal control are approximated as functions of system states in a compact set, which is called \textit{region of interest}, $\Omega$. The \textit{value update} and \textit{policy equation} relations are given as
\begin{equation}
\begin{split}
V^{i+1}(x)&=Q(x)+u^i(x)^TRu^i(x)\\ &+\gamma V^i\big(f(x)+g(x)u^i(x)\big),\forall x\in \Omega,
\label{value_function_VI}
\end{split}
\end{equation}
\begin{equation}
\begin{split}
u^{i}(x)=-\frac{1}{2}\gamma R^{-1}g(x)^T\nabla V^i\big(f(x)+g(x)u^i(x)\big),\forall x\in \Omega,
\label{policy_update_VI}
\end{split}
\end{equation}
The approximation is found by starting from an initial guess for the value function as $V^0(x), \forall x\in \Omega$. Then one uses (\ref{policy_update_VI}) to find $u^0(x),\forall x\in \Omega$. At the next step, (\ref{value_function_VI}) is utilized to find $V^1(x),\forall x\in \Omega$ and so on. The process is repeated until the iterations converge. There are two points regarding this approach. First, it is important to analyze the convergence condition. In other words, what should be the initial guess of $V^0(x), \forall x\in \Omega$, such that the iterations converge to an optimal value? This topic is investigated in \cite{heydari2014revisiting} and it is shown that if $V^0(.)$ is smooth and $0\leq V^0(x)\leq Q(x), \forall x\in \Omega$, then the iterations converge to the optimal solution. The second point regarding the iterations, is solving (\ref{policy_update_VI}). It is seen that on two sides of this equation, $u^i(x)$ appears. Therefore, a system of nonlinear equations with `$m$' equations and `$m$' unknowns should be solved. It is proposed in \cite{heydari2014revisiting} that in order to solve this system of nonlinear equations, one can use another set of iterations with index `$j$' as follows
\begin{equation}
\begin{split}
u^{i,j+1}(x)=-\frac{1}{2}\gamma R^{-1}g(x)^T\nabla V^i\big(f(x)+g(x)u^{i,j}(x)\big),\forall x\in \Omega,
\label{policy_update_iteration_VI}
\end{split}
\end{equation}
Therefore, in order to find $u^i(x)$, one can start with a random initial guess for $u^{i,0}(x)$ and then iterate through (\ref{policy_update_iteration_VI}) until convergence. It is proved in \cite{heydari2014revisiting} that under the following conditions, this iteration will converge:

1) The norm of matrix $R^{-1}$ is small enough.

2) The norm of matrix valued function $g(x)$ is small enough, $\forall x\in \Omega$.

Note that continuous-time state equations are utilized to realize a dynamical system. The presented approach is based on discrete-time dynamics, therefore the system equations are discretized with a sampling time. If the sampling time is small enough, then the two conditions for the convergence of $u^i(x)$ can be satisfied.

In order to implement this approach, the critic and actor neural networks are utilized. For these two networks, linear-in-weight neural networks are used as follows
\begin{equation}
\begin{split}
V(x)\simeq W_c^T\phi(x), \forall x\in \Omega,	
\label{Linear_in_weight_critic}
\end{split}
\end{equation}
\begin{equation}
\begin{split}
u(x)\simeq W_a^T\sigma(x),	\forall x\in \Omega,
\label{Linear_in_weight_actor}
\end{split}
\end{equation}
where $\phi:\mathbb{R}^n \to \mathbb{R}^{n_c}$ and $\sigma:\mathbb{R}^n \to \mathbb{R}^{n_a}$ are the basis functions and positive integers $n_c$ and $n_a$ are the number of neurons in the critic and actor, respectively. Finally, $W_c\in \mathbb{R}^{n_c}$ and $W_a\in \mathbb{R}^{n_a\times m}$ denote the network weights, which will be determined through the training. Note that for each iteration of value function $V^i(x)$, a corresponding critic weight $W_c^i$ is calculated.

Therefore, in order to solve an optimal control problem with ADP and using critic and actor networks, one starts with choosing a large number of random states from the region of interest. These random states will be used for training the critic and actor networks. The training algorithm starts with initializing the value function $V^0(x)$. This initialization should be done for all the randomly selected states. The critic weights, $W_c^0$, can be obtained by using least square method applied on the input-target pairs $(x,V^0(x))$. At the next step, optimal control corresponding to $V^0(x)$ should be calculated. This is done by initializing $u^{0,0}(x)$ and iterating (\ref{policy_update_iteration_VI}) until $u^{0}(x)$ is achieved. After calculating $u^{0}(x)$ for all the randomly selected states, (\ref{value_function_VI}) is used to calculate $V^1(x)$, which leads to the calculation of $W_c^1$. If the conditions of convergence, which are stated above, are met, then the iterations will converge after $\kappa$ iterations. The optimal critic weight $W_c^*$ is calculated from the input-target pair $(x,V^{\kappa}(x))$. The final step is to calculate the optimal actor weights, $W_a^*$, which is obtained from the input-target pair $(x,u^{\kappa}(x))$. In the online control stage, $W_a^*$ is used to find the optimal control $u^*(x)$ at each sampling time. Interested readers are referred to \cite{heydari2014revisiting, heydari2015optimal, zhang2012adaptive} for more details on stability and convergence analysis.

\section{Implementation on PMSM} \label{ADP_PMSM}

\begin{figure}[]
	\centering
	\includegraphics[scale=0.5]{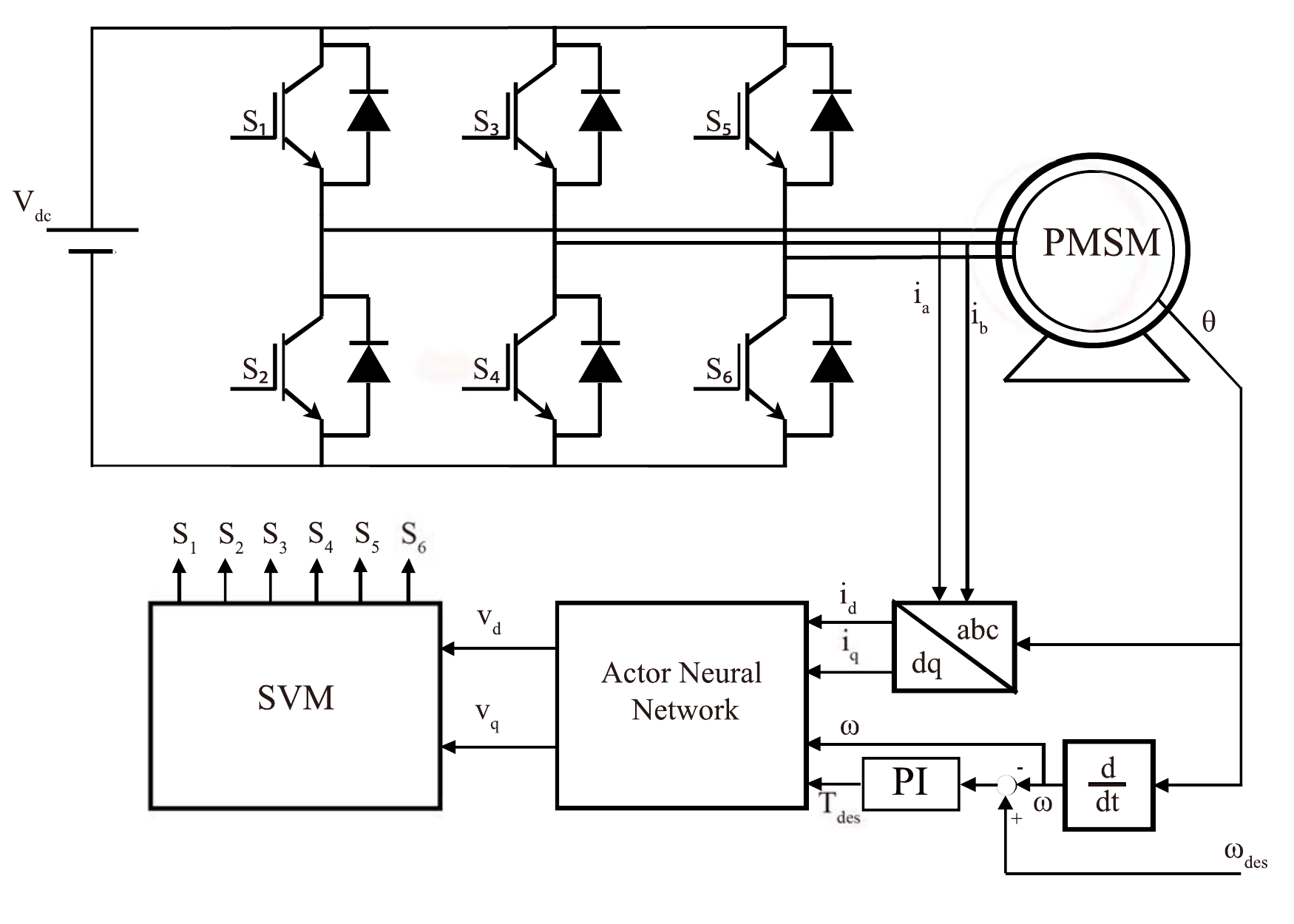}	
	\caption{PMSM control block diagram using actor neural network}
	\label{Fig_block_diagram}
\end{figure}

In this section, the presented ADP-based optimal controller is applied on a Surface Mount-PMSM and its performance is compared with FOC and (PI-based) DTC-SVM, \cite{lai2001new}. Also relative robustness of ADP-based controller under parameter uncertainties is shown by some simulations and it is compared with robustness of FOC and DTC-SVM. Fig. \ref{Fig_block_diagram} shows the control block diagram and the PMSM parameters are given in table \ref{motor_parameters}.

\begin{table}[ht]
	\caption{Motor and Control System Parameters} 
	\centering 
	\begin{tabularx}{0.47\textwidth}{X l}
		\hline\hline
		Parameter & \multicolumn{1}{c}{Value}  \\ [0.5ex] 
		\hline 
		Motor type & Surface Mount-PMSM \\ 
		Pole pairs, $P$ & 5 \\ 
		Permanent magnet flux, $\lambda_m$ & 0.015 $Wb$\\ 
		Stator resistance, $R_s$ & 1.2 $\Omega$ \\ 
		Stator inductance, $L_d=L_q=L_s$ & 0.003 $H$\\ 
		Rated speed, $n_{rated}$ & 3000 $rpm$\\ 
		Rated torque, $\tau_{rated}$ & 0.64 $N.m$\\ 
		Rated output, $P_{out}$ & 0.2 $kW$\\ 
		Rated current, $I_{rated}$ & 2.5 $A(rms)$\\ 
		Maximum speed, $n_{max}$ & 6000 $rpm$\\ 
		Maximum torque, $\tau_{max}$ & 1.91 $N.m$\\ 
		Maximum current, $I_{max}$ & 7 $A(rms)$\\ 
		DC bus voltage, $V_{dc}$ & 100 $V$\\ 
		Rotor moment of inertia, $J$ & 30$\times$10$^{-6}$ $kg.m^2$\\ 
		Sampling time, $T_s$ & 40 $\mu S$\\ 
		\hline\hline
	\end{tabularx}
	\label{motor_parameters} 
\end{table}
\vspace{-.1in}
\subsection{Neural Network Training}\label{NN_training}
The first step in training the neural networks is defining the cost function. As seen in (\ref{cost_function}), the cost function has two terms, $Q(x)$ and $R$. Our objective is tracking torque, while guaranteeing maximum torque per ampere (MTPA) criteria. Motivated by \cite{preindl2013model}, $Q(x)$ can be considered as 
\begin{equation}
\begin{split}
Q(x_k)=K_1\big(\tau_{em}(k)-\tau^*_{em}(k)\big)^2+K_2 i_d(k)^2,
\label{Q_cost_function}
\end{split}
\end{equation}
the constant matrix $R$ is utilized to penalize the control inputs, which are $v_d$ and $v_q$.
\begin{equation}
\begin{split}
R=K_3\begin{bmatrix} 1 & 0 \\
0& 1\end{bmatrix},
\label{R_cost_function}
\end{split}
\end{equation}
where $K_1$, $K_2$, and $K_3\in \mathbb{R}$ are design parameters. Design parameter $K_1$ is the weight penalizing the error between the reference and actual torque, $K_2$ penalizes $i_d$, serving the purpose of maximizing MTPA, and $K_3$ penalizes the control input.

There are two important points to be considered in training. First, as seen in (\ref{Q_cost_function}), the objective is tracking the reference torque, therefore this value is needed to find the control signals. In order to consider the effect of reference torque on the control signal, the inputs to the critic and actor are augmented, by adding the reference torque to system states. This way the controller can ``learn'' how to react when the reference torque changes. Secondly, according to equations (\ref{value_function_VI}) and (\ref{policy_update_VI}), the right hand side needs the next state $x_{k+1}$. It is observed in (\ref{motor_dynamics}) that in order to find the next state, the speed $\omega_m$ is needed. Therefore, $\omega_m$ is added to the inputs of critic and actor as an ``exogenous'' input. Therefore, the optimal value function and optimal control are functions of discrete-time states ($x=[i_d,i_q]^T$), reference torque ($\tau^*_{em}$), and motor mechanical speed ($\omega_m$) . Equations (\ref{value_function_VI}), (\ref{policy_update_VI}), (\ref{Linear_in_weight_critic}), (\ref{policy_update_iteration_VI}), and (\ref{Linear_in_weight_actor}) are modified as

\begin{equation}
\begin{split}
V^{i+1}(x,\tau^*_{em},\omega_m)&=Q(x,\tau^*_{em})\\ &+u^i(x,\tau^*_{em},\omega_m)^TRu^i(x,\tau^*_{em},\omega_m)\\ &+\gamma V^i\big(f(x,\omega_m)+gu^i(x,\tau^*_{em},\omega_m)\big),\\
&\forall [x^T,\tau^*_{em},\omega_m]^T\in \Omega,
\label{value_function_VI_augmented}
\end{split}
\end{equation}
\begin{equation}
\begin{split}
u^{i}(x,\tau^*_{em},\omega_m)&=-\frac{1}{2}\gamma R^{-1}g^T\nabla V^i\big(f(x,\omega_m)\\ &+gu^i(x,\tau^*_{em},\omega_m)\big),
\forall [x^T,\tau^*_{em},\omega_m]^T\in \Omega,
\label{policy_update_VI_augmented}
\end{split}
\end{equation}
\begin{equation}
\begin{split}
u^{i,j+1}(x,\tau^*_{em},\omega_m)&=-\frac{1}{2}\gamma R^{-1}g(x)^T\nabla V^i\big(f(x,\omega_m)\\ &+gu^{i,j}(x,\tau^*_{em},\omega_m)\big),\forall [x^T,\tau^*_{em},\omega_m]^T\in \Omega,
\label{policy_update_iteration_VI_augmented}
\end{split}
\end{equation}
\begin{equation}
\begin{split}
V(x,\tau^*_{em},\omega_m)\simeq W_c^T\phi(x,\tau^*_{em},\omega_m), \forall [x^T,\tau^*_{em},\omega_m]^T\in \Omega,	
\label{Linear_in_weight_critic_augmented}
\end{split}
\end{equation}
\begin{equation}
\begin{split}
u(x,\tau^*_{em},\omega_m)\simeq W_a^T\sigma(x,\tau^*_{em},\omega_m), \forall [x^T,\tau^*_{em},\omega_m]^T\in \Omega,
\label{Linear_in_weight_actor_augmented}
\end{split}
\end{equation}

As mentioned in Section \ref{Optimal_control_ADP}, the region of interest ($\Omega$) for the states should be selected for the training. If elements of the input vector to the networks assume values within comparable ranges, the approximations will have better results, \cite{krose1993introduction}. Therefore, the maximum values of current, speed and torque are used to normalize the states
\begin{equation}
\begin{split}
&i_d=I_{max}\tilde{i}_d, i_q=I_{max}\tilde{i}_q, \\
&\tau^*_{em}=\tau_{max}\tilde{\tau}^*_{em}, \omega_m=\omega_{m_{max}}\tilde{\omega}_m, \\
\label{state_normalization}
\end{split}
\end{equation}
where the normalized quantities are denoted by ` $\mathtt{\sim}$ ' notation. Therefore, the region of interest is selected as
\begin{equation}
\begin{split}
\Omega=&\{[\tilde{i}_d, \tilde{i}_q, \tilde{\omega}_m, \tilde{\tau}^*_{em} ]^T \in \mathbb{R}^4 \\
& : -1.5 \leq \tilde{i}_d, \tilde{i}_q, \tilde{\omega}_m, \tilde{\tau}^*_{em} \leq 1.5 \}.
\label{region_of_interest}
\end{split}
\end{equation}
This selection makes sure that the maximum value of the normalized states, which is 1, is well inside the training set, therefore better generalization is achieved near the boundaries of the variation of the states.

\begin{figure}[!t]
	\centering
	\includegraphics[scale=0.58]{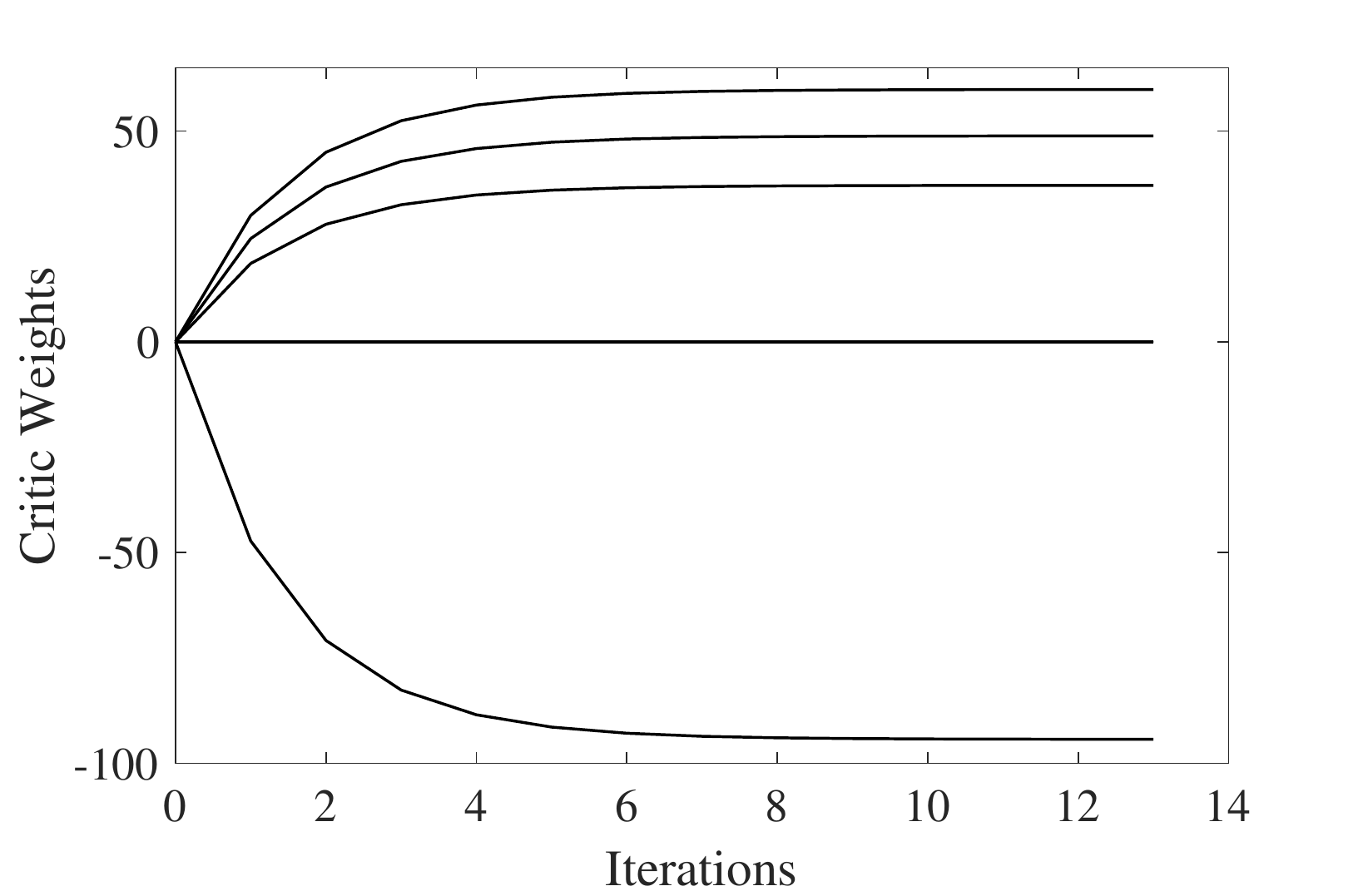}	
	\caption{History of critic weights during learning iterations}
	\label{critic_weights}
\end{figure}

Since linear-in-weight neural networks are used for critic and actor, basis functions $\phi$ and $\sigma$ should be chosen in (\ref{Linear_in_weight_critic}) and (\ref{Linear_in_weight_actor}). Motivated by Weierstras approximation theorem,  the basis functions are chosen as
\begin{equation}
\begin{split}
&\phi(\eta)=[1,\eta^T,(\eta \otimes \eta)^T,\big(\eta \otimes (\eta \otimes \eta) \big)^T]^T,\\
&\sigma(\eta)=[1,\eta^T,(\eta \otimes \eta)^T]^T,\\
&\eta=[x,\tau^*_{em},\omega_m],
\end{split}
\end{equation}
where $\eta \otimes \eta$, is the non-repeating polynomials built from multiplying elements of vector $\eta$ by those of $\eta$. Since $\eta \in \mathbb{R}^4$, there is 35 neurons for the critic network and 15 neurons for the actor network. These basis function are design parameters, therefore, one can make another choice. The number of 10000 random states are selected from the region of interest to do the batch learning algorithm \cite{heydari2014revisiting}. The training is done by $K_1=30$, $K_2=0.5$, $K_3=100$, and $\gamma=0.5$. Fig. \ref{critic_weights} shows the weights converging after 12 iterations, which took almost 45 seconds on a desktop computer with Intel Core i-5-6500, 3.2 GHz processor, and 16 GB of RAM, running MATLAB 2018a.

After training and obtaining the weights, the actor weights are calculated using the least square method. The history of critic weights and optimal actor weights can be found in \cite{Motor_training_files}. Thanks to the obtained actor weights, the only computations required in online stage of the control is to calculate $W^*_a \sigma(\eta)$, which will lead to control signals $v_d$ and $v_q$. These voltages are then applied on the motor by the SVM block.

The following pseudocodes show the training and implementation of this approach.

\begin{algorithm}
	\DontPrintSemicolon
	\SetAlgoLined
	\BlankLine
	\SetKwInOut{Initialization}{Initialization}
	\Initialization{Choose the following:\\
		Scaling factors according to (\ref{state_normalization}).\\
		$\bar{n}$ random state $\eta^{[q]}\in\Omega, \forall q \in \{1,2,...,\bar{n}\}$\\
		Basis functions $\phi(\eta)$, $\sigma(\eta)$.\\
		$Q$ and $R$ according to (\ref{Q_cost_function}), (\ref{R_cost_function}).\\
		$\beta_v$ and $\beta_u$ as convergence tolerances.
	}
	
	Set $i=0$.
	
	Initialize $V^i(\eta^{[q]})=0,\forall q \in \{1,2,...,\bar{n}\}$.
	
	Find $W_c^i$ from input-target pair $(\eta^{[q]},V^i(\eta^{[q]}))$.
	
	Set $j=0$.
	
	Initialize $u^{i,j}(\eta^{[q]})$ with a random value.
	
	Calculate $u^{i,j+1}(\eta^{[q]})$ from (\ref{policy_update_iteration_VI_augmented}).
	
	If $||u^{i,j+1}(\eta^{[q]})-u^{i,j}(\eta^{[q]})||<\beta_u$, proceed to the next step, otherwise set $j=j+1$ and go back to step 6.
	
	Calculate $V^{i+1}(\eta^{[q]}),\forall q \in \{1,2,...,\bar{n}\}$ from (\ref{value_function_VI_augmented}).
	
	If $||V^{i+1}(\eta^{[q]})-V^{i}(\eta^{[q]})||<\beta_v$, proceed to the next step, otherwise go back to step 3.
	
	Find $W_a^*$ from input-target pair $(\eta^{[q]},u^i(\eta^{[q]}))$.

	\caption{The pseudocode for the offline training of actor and critic weights}
	\label{Pseudocode_for_training}
\end{algorithm}

\begin{algorithm}
	\DontPrintSemicolon
	\SetAlgoLined
	\BlankLine
	\SetKwInOut{Initialization}{Initialization}
	\Initialization{Save the optimal actor weights $W_a^*$.\\
		Save actor basis function $\sigma(\eta)$.\\
	}
	
	Read motor currents $i_a$ and $i_b$.
	
	Read rotor position $\theta_m$ and calculate the rotor speed $\omega_m$.
	
	Calculate d-q currents from $i_a$ and $i_b$ using (\ref{PK_transform}).
	
	Pass the error between desired speed and actual speed to a PI controller to get the reference torque.
	
	Use (\ref{state_normalization}) to normalize the currents, speed and torque.
	
	Calculate $v_d$ and $v_q$ using $W_a^*\sigma(\eta)$
	
	Feed $v_d$ and $v_q$ to the SVM block to generate the appropriate switching.
	
	Go back to step 1.

	\caption{The pseudocode for the online implementation of motor control}
	\label{Pseudocode_for_implementation}
\end{algorithm}

\subsection{Comparative simulation}\label{simulation}
\begin{figure}[!t]
	\centering
	\includegraphics[scale=0.48]{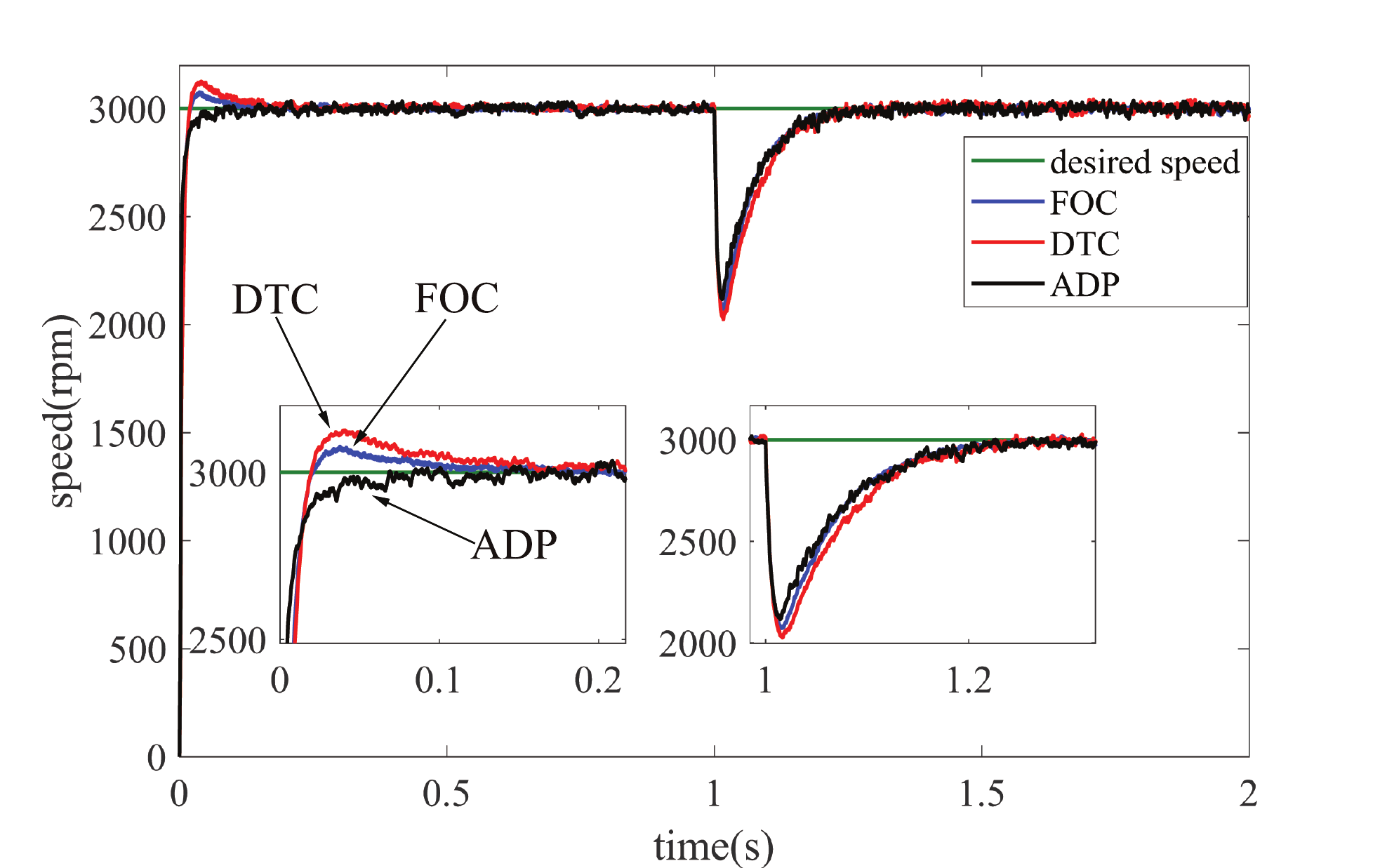}	
	\caption{Speed response simulation of three controllers when a load torque is applied on the motor at rated speed}
	\label{speed_test_sudden_load}
\end{figure}

In this section, the calculated actor weights in previous section are used to do simulations on the motor and compare the results with FOC and PI-based DTC-SVM. The PI gains of the speed loop are selected to be identical for all three control algorithms, so that the difference in the performance is only due to the difference in their torque tracking capability. In the first simulation, the motor shaft rotates at its nominal speed, i.e., 3000 rpm, and a load torque is applied on it. Fig. \ref{speed_test_sudden_load} shows the motor speed with three control algorithms, when a load torque of 0.6 $N.m$ is applied on the motor at $t=1s$. It is seen that the results for three controllers are very close but ADP is showing a slightly better performance, both in reaching the desired speed from standstill and after the load torque is applied. These two parts of the figure are magnified so that the transient response of each algorithm can be seen clearly.

\begin{figure}[!t]
	\centering
	\includegraphics[scale=0.47]{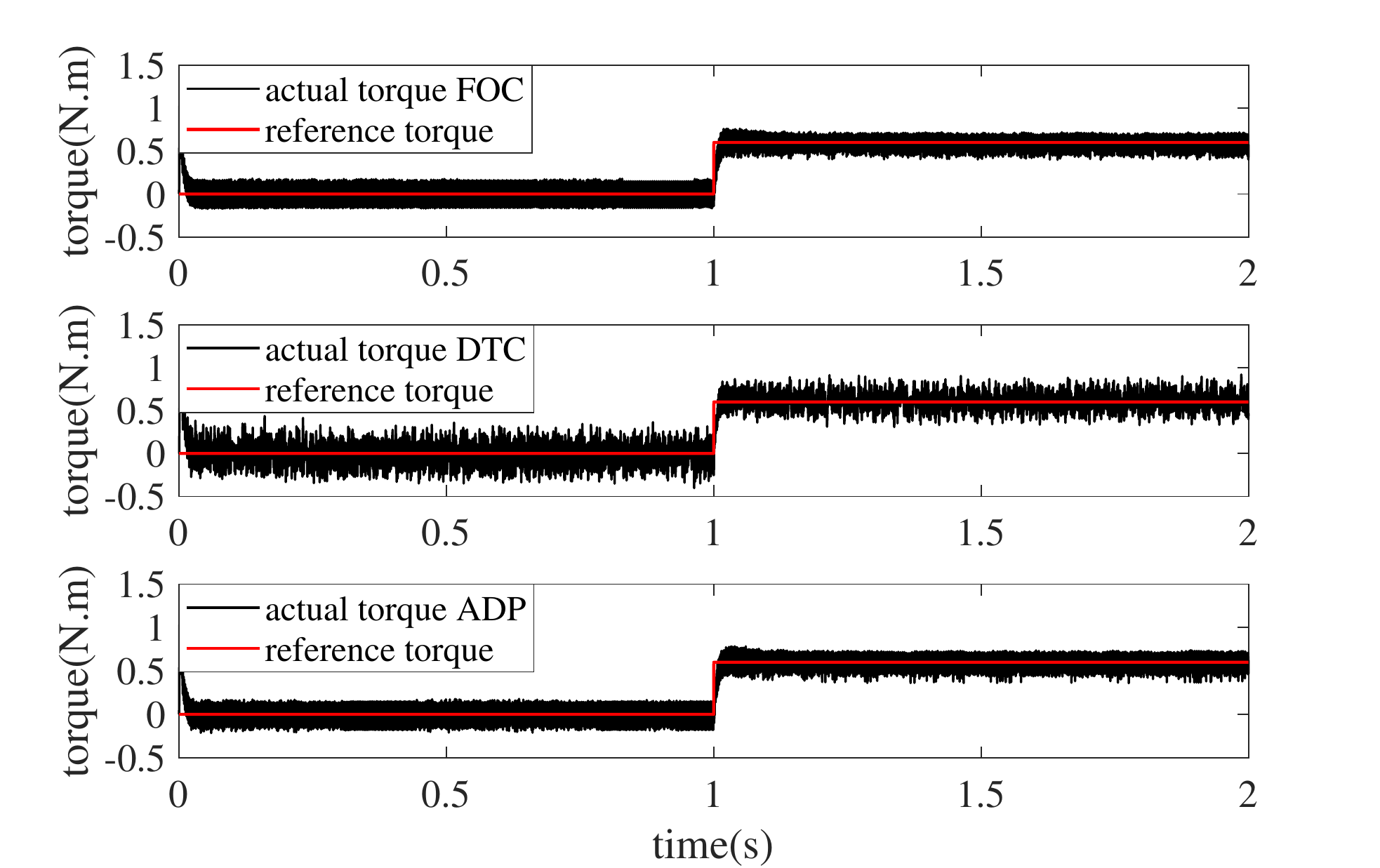}	
	\caption{Torque response simulation of three controllers when a load torque is applied on the motor at the rated speed of 3000 rpm}
	\label{torque_response}
\end{figure}

Fig. \ref{torque_response} shows the generated torque of motor under three control algorithms. It is seen that all three approaches have a fast torque dynamic, however the DTC-SVM algorithm has more torque ripple compared to the other two approaches. Table. \ref{sudden_torque_ITAE}, shows the calculated integral time absolute error (ITAE) for the three algorithms. It is seen that the ITAE of ADP is slightly better than those of FOC and DTC-SVM.

\begin{table}[ht]
	\caption{ITAE of motor torque when a load torque of 0.6 $N.m$ is applied\\
		after one second on the motor at the rated speed of 3000 rpm} 
	\centering 
	\begin{tabularx}{0.47\textwidth}{X l}
		\hline\hline
		Control Algorithm & {Torque ITAE}   \\ [0.5ex] 
		\hline 
		FOC & 0.0251 \\ 
		DTC-SVM & 0.0287 \\ 
		ADP & 0.0245\\  
		\hline\hline
	\end{tabularx}
	\label{sudden_torque_ITAE} 
\end{table}

\begin{figure}[!t]
	\centering
	\includegraphics[scale=0.48]{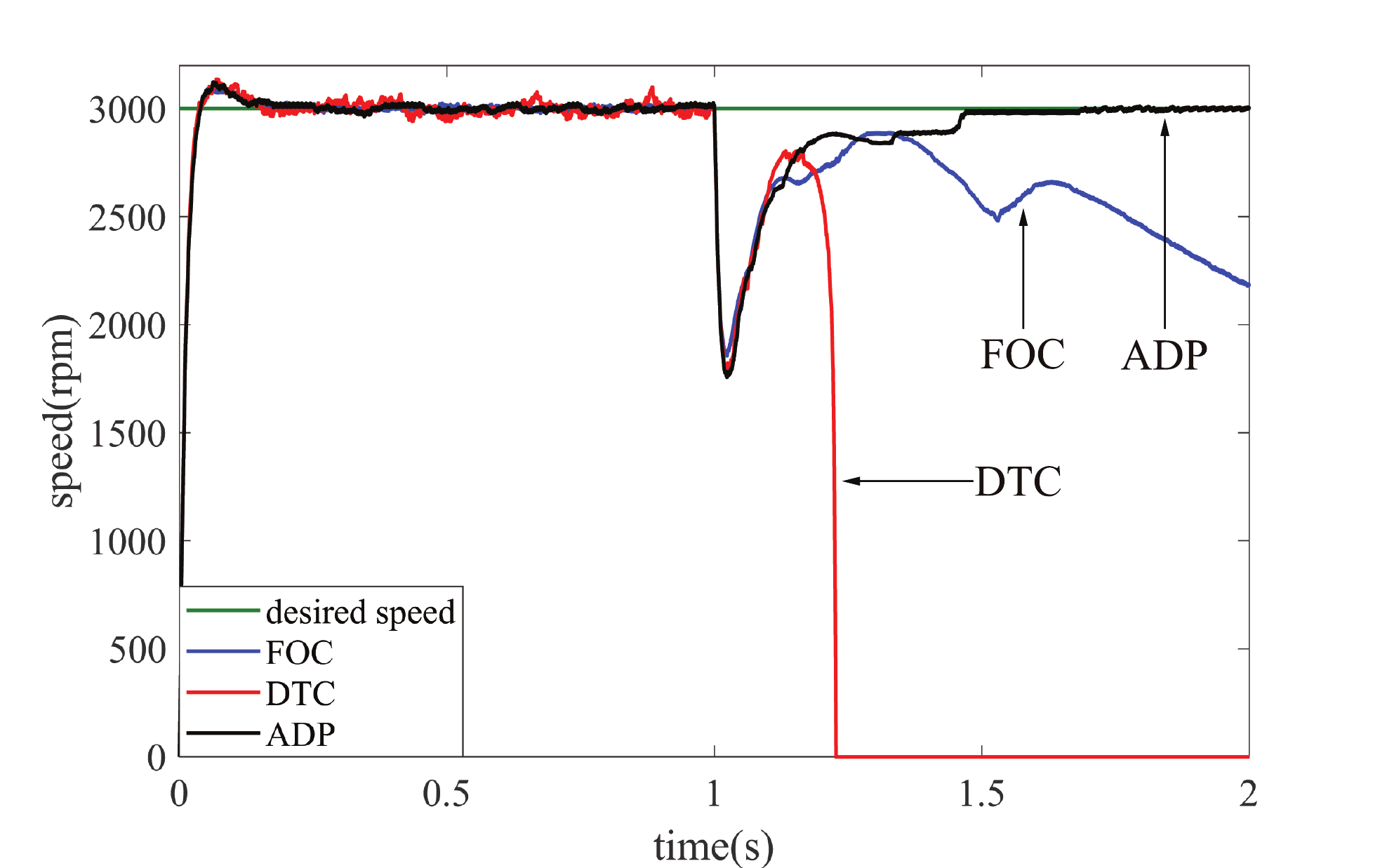}	
	\caption{Speed response simulation of three controllers when a load torque is applied on the motor at rated speed with parameter uncertainties}
	\label{speed_test_sudden_load_uncertain}
\end{figure}

The second simulation considers parameter uncertainties in the motor. In order to evaluate the performance of the algorithms in handling uncertainties, the controllers are tuned using nominal values, however in online control, the parameters of the motor with uncertainties are used for simulation. In other words, for instance, in ADP, all the trainings are done with the nominal parameters of Table \ref{motor_parameters}, but the motor simulations are done with parameters of Table \ref{Uncertain parameters in simulation}. This way the controller has not learned how to react to new parameters. This test can be used to demonstrate the relative robustness of each algorithm.
\begin{table}[ht]
	\caption{Motor Parameters with uncertainties} 
	\centering 
	\begin{tabularx}{0.47\textwidth}{X l}
		\hline\hline
		Parameter & \multicolumn{1}{c}{Value}  \\ [0.5ex] 
		\hline 
		Permanent magnet flux, $\hat{\lambda}_m$ & 0.012 $Wb$\\ 
		Stator resistance, $\hat{R}_s$ & 5.7 $\Omega$ \\ 
		Stator inductance, $\hat{L}_s$ & 0.001 $H$\\   
		Rotor moment of inertia, $\hat{J}$ & 40$\times$10$^{-6}$ $kg.m^2$\\ 
		\hline\hline
	\end{tabularx}
	\label{Uncertain parameters in simulation} 
\end{table}
Again the motor response is tested when it is rotating at the rated speed and a load torque of 0.6 $N.m$ is applied at $t=1s$. It is seen in Fig. \ref{speed_test_sudden_load_uncertain} that before the load is applied, all three controllers can track the desired speed, however when the load is applied at $t=1s$, only ADP is capable of tracking the reference. A poor performance for DTC-SVM was expected, given its model dependency. However, FOC is not explicitly dependent on the parameters like stator reluctance and resistance as well as permanent magnet flux. Therefore, one may expect an acceptable performance from FOC under such modeling imperfections. But, as see in this figure, FOC also fails, once the load torque is applied. The reason for the poor performance of FOC approach is that, the current loop PI gains are tuned based on the motor nominal parameters. Therefore, when the motor parameters have changed to the parameters in Table.\ref{Uncertain parameters in simulation}, the PI controller of FOC cannot deal with the uncertainties.

In conclusion of the simulation study, one sees that ADP provides significantly better results in dealing with parameter uncertainties. However, It should be noted that the ADP controller is not analytically designed to be robust to parameter uncertainties and it is not claimed here that ADP is robust to structured and unstructured uncertainties. However, because of the feedback nature of the controller and its learning capability, it is seen in simulations that uncertainties are managed to some extent. If the uncertainties are more significant, the desirable performance is not guaranteed.

\section{Experimental results} \label{ADP_PMSM_experiment}
\begin{figure}[!t]
	\centering
	\includegraphics[scale=0.3]{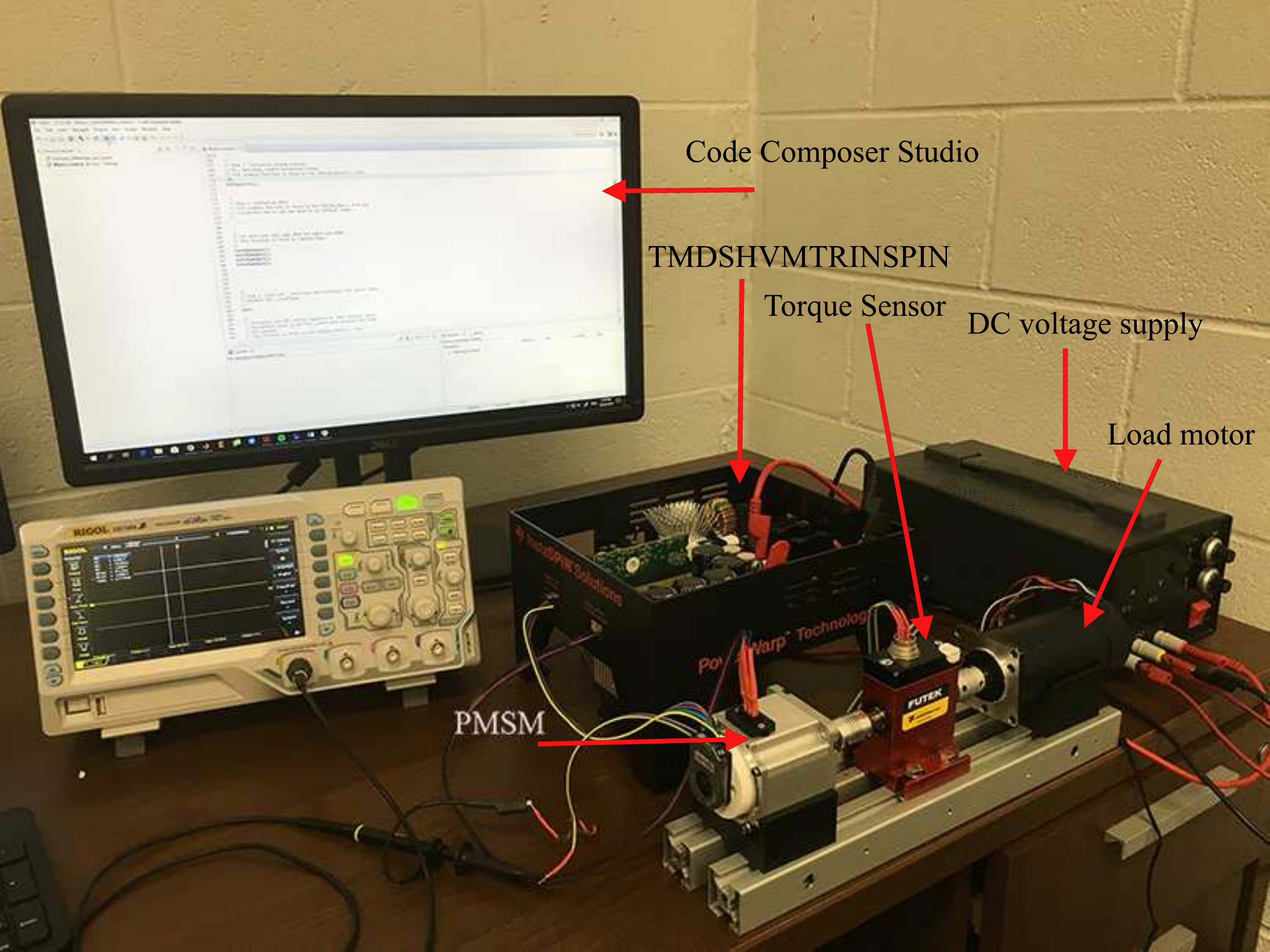}	
	\caption{PMSM control experimental setup}
	\label{Test_setup}
\end{figure}

In this section, the proposed ADP approach is implemented on a hardware testbed and its performance is compared with FOC and DTC-SVM, experimentally.  Fig. \ref{Test_setup} shows the experimental setup. The PMSM model parameters are the same as Table. \ref{motor_parameters}. A BLDC motor is used as a load to apply external step torque on PMSM motor shaft. The Texas Instrument development kit TMDSHVMTRINSPIN, with a microcontroller TMS320F28069M is used to implement controllers. This microcontroller has a 90 MHz clock, 16 PWM channels and 16 channels of 12-Bit analog to digital converters (ADC). It is also capable of doing floating point operations very fast, which makes it a suitable processor for this application. The experiments in this section are done under both nominal motor parameters and parameter uncertainties. 

\subsection{Comparative experiment under nominal condition}\label{experiment_nominal}
\begin{figure}[!t]
	\centering
	\includegraphics[scale=0.48]{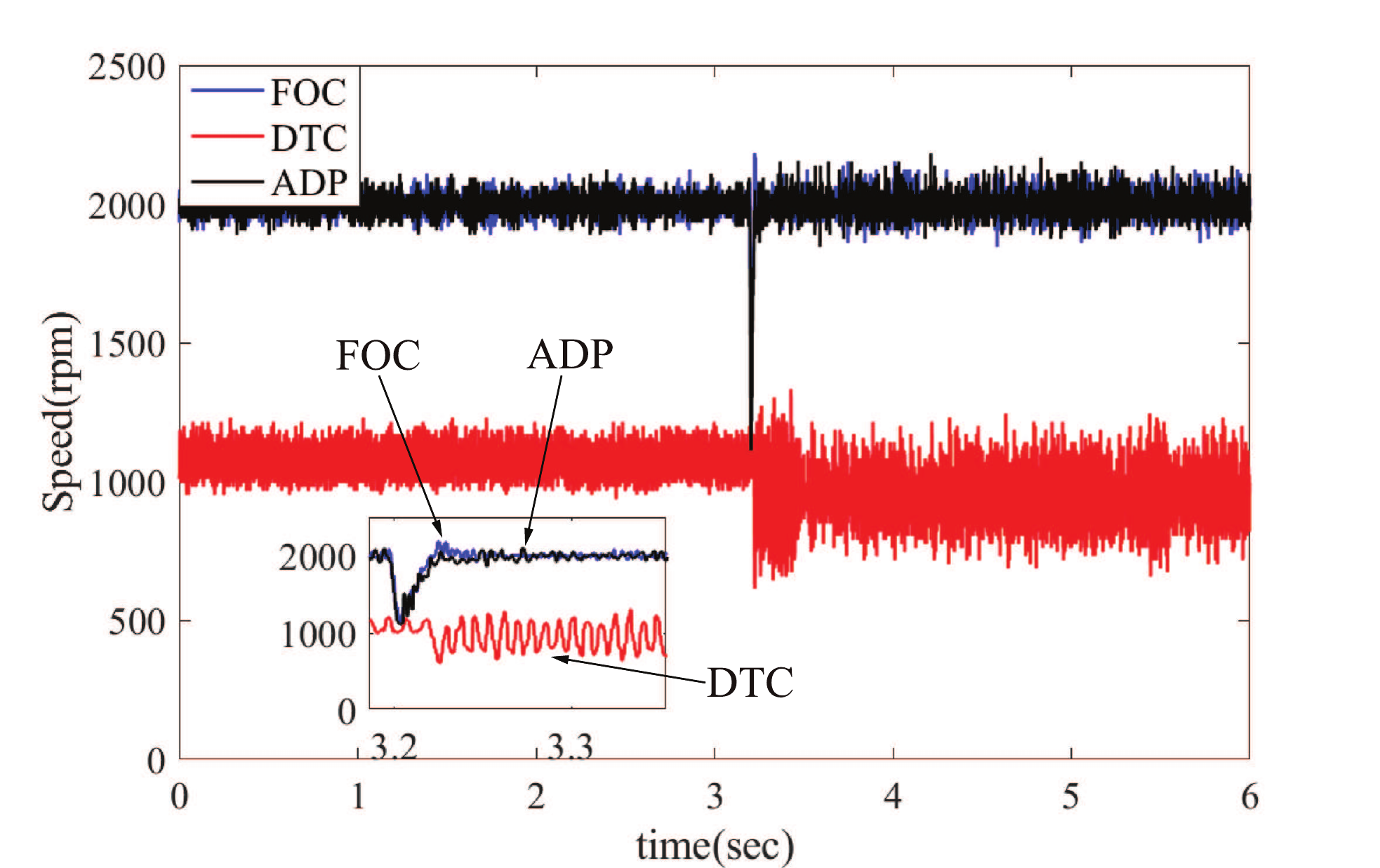}	
	\caption{Speed response experiment of three controllers when a load of 0.7 $N.m$ is applied at 2000 $rpm$}
	\label{speed_test_sudden_load_EXP}
\end{figure}

The same actor weights, which were obtained in simulation are used for the experiments in this section. It should be noted that, in order to have a fair comparison, the PI gains of the speed loop are considered the same for all three algorithms. In the first experiment, the motor is running at 2000$rpm$ and a step load of 0.7$N.m$ is applied on the motor at $t=2.3s$. Fig. \ref{speed_test_sudden_load_EXP}, shows the response of each algorithm. Two sections of the figure are magnified to show the transient response and steady state response of the speed. It is observed that ADP and DTC-SVM are faster than FOC in reaching the desired speed after the torque is applied. However, DTC-SVM shows more ripples during steady state.

\begin{figure}[!t]
	\centering
	\includegraphics[scale=0.48]{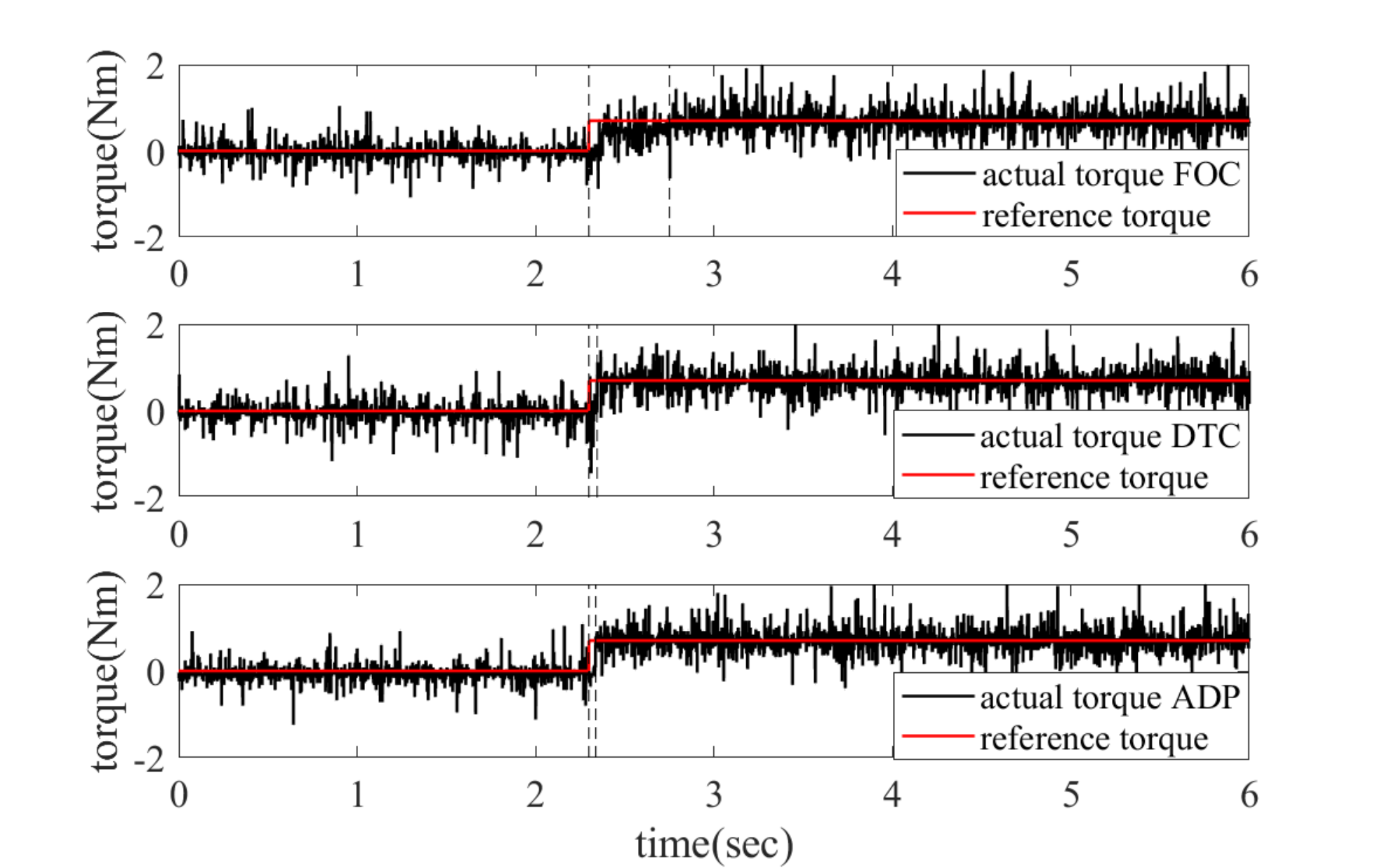}	
	\caption{Torque response experiment of three controllers when a load of 0.7 $N.m$ is applied at 2000 $rpm$}
	\label{torque_response_EXP}
\end{figure}

For the above experiment, the torque is also measured using a torque sensor, namely, FUTEK FSH02564 . Fig. \ref{torque_response_EXP} shows the generated torque for each algorithm. The vertical dashed lines show the time duration of the transient response of the torque. This time duration for FOC, DTC-SVM, and ADP is 0.4521$s$, 0.0462$s$, and 0.039$s$, respectively.  It is seen that the torque response dynamics for FOC is slower than ADP and DTC-SVM.

Table \ref{sudden_torque_ITAE_EXP} shows the ITAE of speed and torque for each algorithms, as points at a slightly better performance for ADP.  Therefore, not only the dynamic response of ADP is fast, but also its torque ripples are comparable with FOC. So both transient and steady state responses are satisfactory with ADP. The cost function, defined in (\ref{cost_function}), with $Q$ and $R$ as (\ref{Q_cost_function}) and (\ref{R_cost_function}) is calculated for FOC, DTC-SVM and ADP as $0.54$, $0.56$, $0.51$, respectively. Although the costs are close, ADP has the lowest cost among the all three approaches.

\begin{table}[ht]
	\caption{ITAE of motor torque and speed when a load torque of 0.7 $N.m$ is\\
		applied at 2000 rpm} 
	\centering 
	\begin{tabularx}{0.47\textwidth}{X l l}
		\hline\hline
		Control Algorithm  & {Speed ITAE} & {Torque ITAE} \\ [0.5ex] 
		\hline 
		FOC  & 146.6808 & 1.1994 \\ 
		DTC-SVM  & 268.0128 & 1.2143 \\ 
		ADP  & 131.0423 & 1.1988\\  
		\hline\hline
	\end{tabularx}
	\label{sudden_torque_ITAE_EXP} 
\end{table}

\begin{figure}[!t]
	\centering
	\includegraphics[scale=0.33]{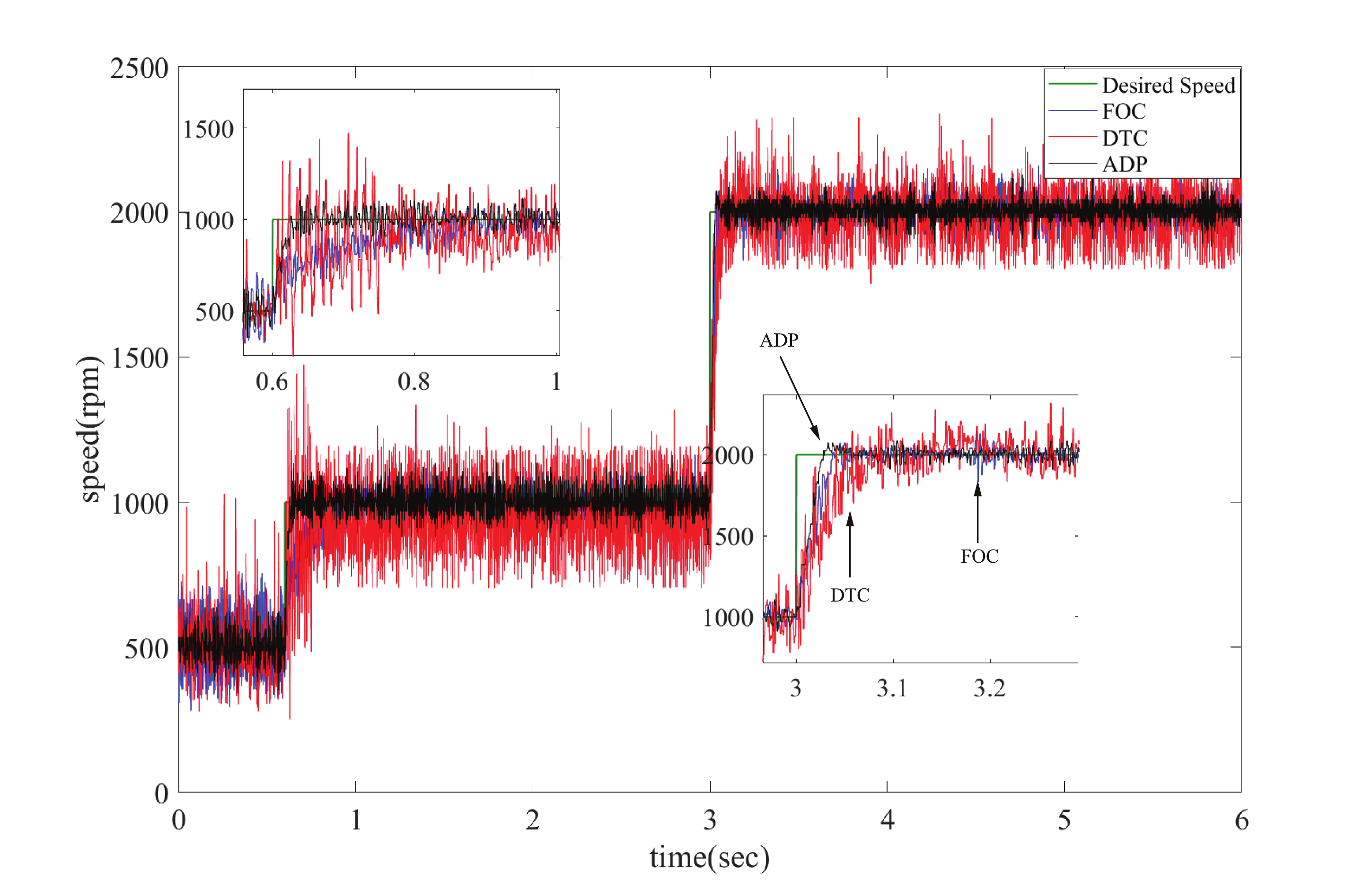}	
	\caption{Speed response experiment of three controllers when motor is under the load of 0.7 $N.m$}
	\label{speed_test_full_load_EXP}
\end{figure}

In the next experiment, the capability of each algorithm in tracking a desired speed under full load of 0.7 $N.m$ is analyzed. The desired speed is given by
\begin{equation}
\begin{cases}
\omega_{ref}=500 rpm& \text{if } 0s\leq t\leq 0.5s\\
\omega_{ref}=1000 rpm& \text{if } 0.5s\leq t\leq 3s\\  
\omega_{ref}=2000 rpm& \text{if } 3s\leq t\leq 6s
\end{cases}
\label{desired_speed_condition_EXP}
\end{equation}
As seen in Fig. \ref{speed_test_full_load_EXP}, FOC and DTC-SVM are not able to follow the desired speed in both directions fast enough. Also the motor speed under DTC-SVM shows high ripples. It is observed that ADP has a fast transient response and a more smooth steady state response in tracking the desired speed.

\subsection{Comparative experiment with parameter variations}\label{experiment_uncertainty}
\begin{figure}[!t]
	\centering
	\includegraphics[scale=0.48]{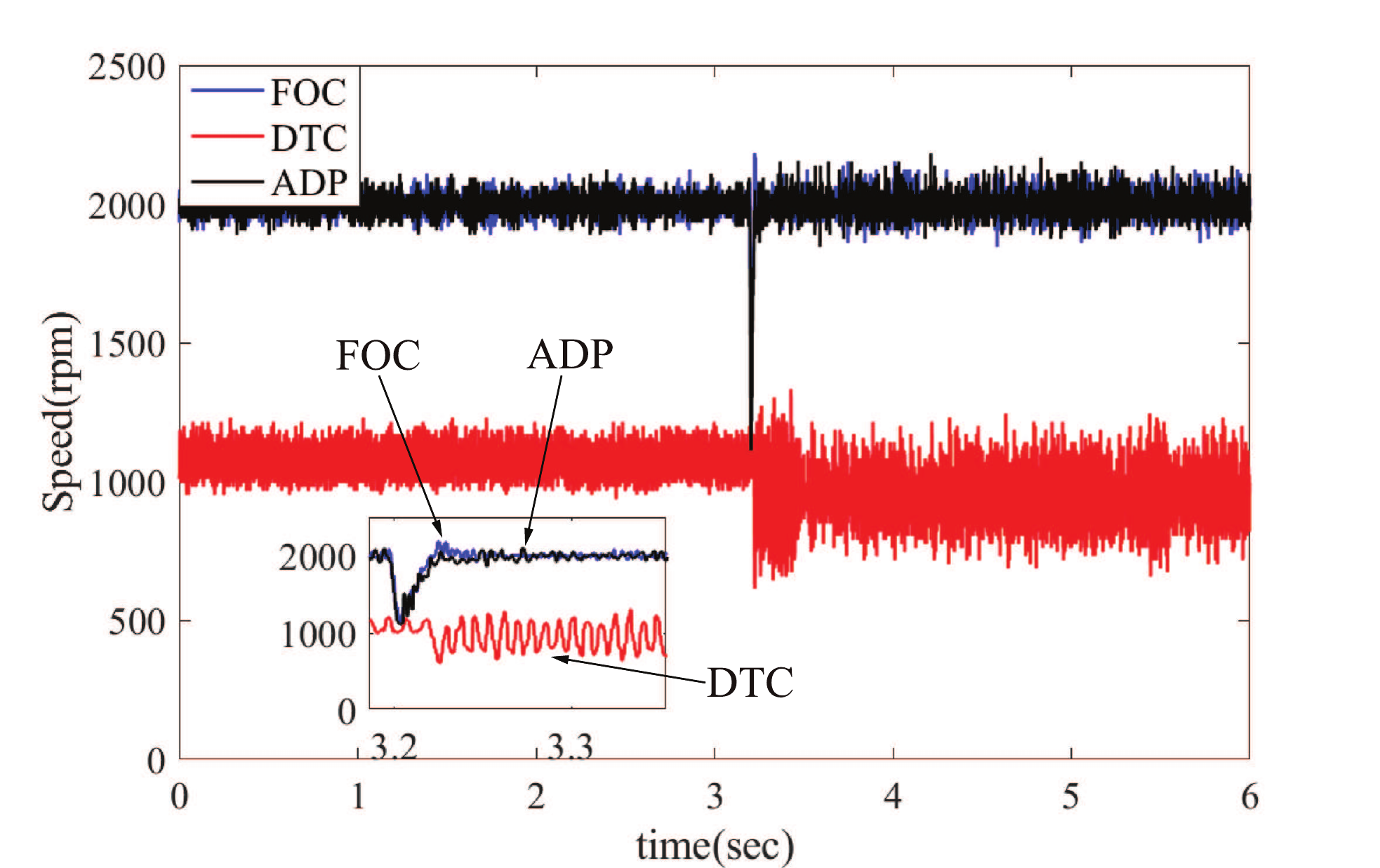}	
	\caption{Speed response experiment of three controllers under parameter uncertainties when a step load of 0.7 $N.m$ is applied at 2000 $rpm$}
	\label{speed_test_sudden_load_uncertain_EXP}
\end{figure}

In order to show the response of each control algorithm in the presence of parameter uncertainties, it is assumed that the information about the motor parameters are incorrect. These incorrect parameters are considered as $R_s=3.6 \Omega$, $L_s=0.001H$, and $\lambda_m=0.005 Wb$. Therefore, controllers are designed based on the incorrect parameter values. However, the actual values are used for the experiment. For ADP, these perturbed parameters are used to obtain the new actor weights. Fig. \ref{speed_test_sudden_load_uncertain_EXP} shows the speed response of each algorithm when a load of 0.7 $N.m$ is applied at $t=3.2s$. It is seen that because of the parameter uncertainties, DTC-SVM can not track the desired speed either before or after the load is applied. This was expected since DTC-SVM is dependent on motor parameters. However, ADP and FOC still show good performances.

\begin{figure}[!t]
	\centering
	\includegraphics[scale=0.48]{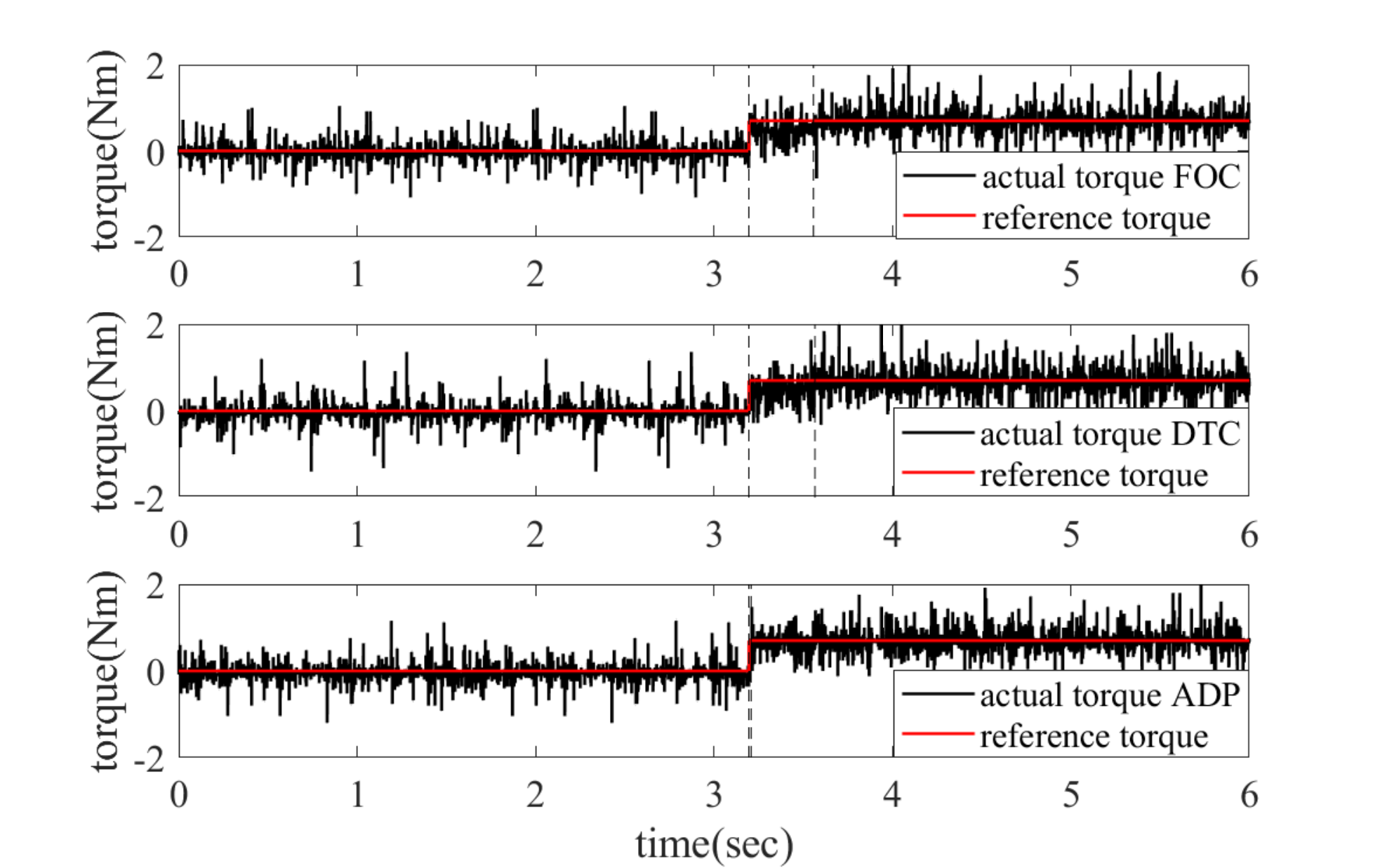}	
	\caption{torque response experiment of three controllers under parameter uncertainties when a step load of 0.7 $N.m$ is applied at 2000 $rpm$}
	\label{torque_response_uncertain_EXP}
\end{figure}

The measured torque is also shown in Fig. \ref{torque_response_uncertain_EXP}.  The vertical dashed lines show the time duration of the transient response of the torque. This time duration for FOC, DTC-SVM, and ADP is 0.3621$s$, 0.3702$s$, and 0.012$s$, respectively. It is observed that, the torque dynamics of DTC-SVM, which was fast with nominal parameters, is now slower under parameter uncertainties. However, the time duration for FOC and ADP has not changed significantly.

Table \ref{sudden_torque_ITAE_EXP_uncertain} shows the ITAE of speed and torque for each algorithms. The cost function for FOC, DTC-SVM and ADP as $0.5418$, $0.5482$, $0.5413$, respectively. 

\begin{table}[ht]
	\caption{ITAE of motor torque and speed under parameter uncertainties\\ when a load torque of 0.7 $N.m$ is
		applied at 2000 rpm } 
	\centering 
	\begin{tabularx}{0.47\textwidth}{X l l}
		\hline\hline
		Control Algorithm  & {Speed ITAE} & {Torque ITAE} \\ [0.5ex] 
		\hline 
		FOC  & 133.5428 & 1.1994 \\ 
		DTC-SVM  & 2419 & 1.3943 \\ 
		ADP  & 132.0423 & 1.1988\\  
		\hline\hline
	\end{tabularx}
	\label{sudden_torque_ITAE_EXP_uncertain} 
\end{table}

\begin{figure}[!t]
	\centering
	\includegraphics[scale=0.33]{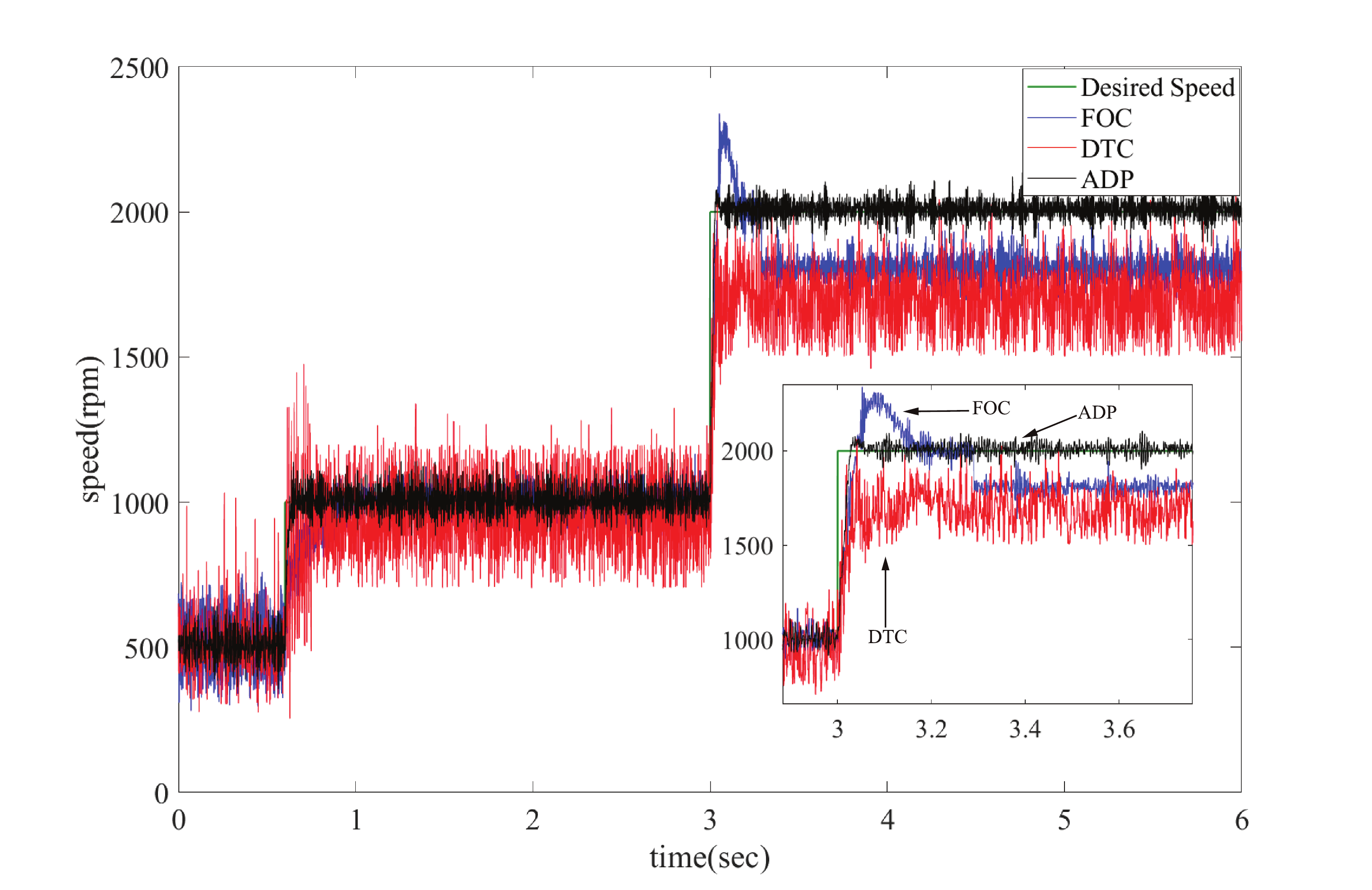}	
	\caption{Speed response experiment of three controllers under parameter uncertainties when motor is under the load of 0.7 $N.m$}
	\label{speed_test_full_load_uncertain_EXP}
\end{figure}

The speed tracking of three algorithms when the motor is under the load of 0.7 $N.m$ is also investigated. The reference speed is as (\ref{desired_speed_condition_EXP}). Fig. \ref{speed_test_full_load_uncertain_EXP} shows superior performance of ADP compared to FOC and DTC-SVM under parameter uncertainties. It is observed that neither FOC nor DTC-SVM can track the desired speed in this case.

All in all, the experiments have also verified the results demonstrated in the simulations. These experiments justifies the relative robustness of ADP to parameter uncertainties. However, as stated before, if the parameter uncertainties are more significant, there is no guarantee to achieve a desired performance.

\section{Conclusion} \label{Conclusion}
An ADP-based control approach is proposed in this paper to achieve fast and accurate torque control in a PMSM. The critic and actor weights are trained once offline and the calculated optimal actor weights are utilized in online control. The simulations and experimental results show that while the advantages of ADP over popular practices, namely, FOC and DTC-SVM are small in the case of \textit{perfect modeling} of the dynamics, the improvements are significant under the case of having \textit{modeling uncertainties} which is a \textit{reality} in any application. In other words, as seen in the experimental results, the performance for both FOC and ADP are similar under nominal conditions. However, when parameter variations are considered, ADP has a better performance compared to FOC and DTC-SVM.
 
\bibliography{Manuscript}
\bibliographystyle{ieeetr}

\end{document}